\documentclass[letterpaper, 10 pt, conference]{ieeeconf}
\usepackage[utf8]{inputenc}
\IEEEoverridecommandlockouts

\let\labelindent\relax
\usepackage{enumitem}
\usepackage{algorithm}

\usepackage{algpseudocode}
\usepackage{amsmath, amssymb}

\newtheorem{theorem}{Theorem}
\newtheorem{lemma}{Lemma}

\newcommand{\qedsymbol}{\mbox{ }\hfill$\square$}

\usepackage{booktabs}                 							% nicer tables, nice!
\usepackage{caption}
\usepackage{fancyhdr}                   						% Nice and shiny headers
\usepackage{graphicx}                   						% For images
\usepackage[table]{xcolor}
\usepackage{arydshln}
\usepackage{tikz}
\usetikzlibrary{arrows.meta}
\usepackage[greek, british]{babel}     							% Use british hyphenation stuff and the Greek Euro symbol

\newcommand{\bc}{\cellcolor{blue!10}}
\newcommand{\rc}{\cellcolor{red!10}}
\newcommand{\gc}{\cellcolor{green!10}}

\title{\LARGE Model reduction for linear parameter-varying \\systems through parameter projection}

\author{Sil Schouten, Daming Lou and Siep Weiland $^{1}$
\thanks{$^{*}$ Accepted by $58^{\text{th}}$ IEEE Conference on Decision and Control.}
\thanks{$^{1}$ Department of Electrical Engineering, Eindhoven University of Technology, Eindhoven, The Netherlands. Emails: {\tt\small \{S.Schouten, D.Lou, S.Weiland\}@tue.nl}
}
}

\begin{document}
\maketitle
\thispagestyle{plain}
\pagestyle{plain}
\begin{abstract}
%This paper deals with parameter reduction of linear parameter-varying (LPV) systems as a means of reducing complex models. %First, a number of system norms for parameter-varying systems are introduced. These norms on LPV systems are used to evaluate the performance of the reduction methods.
%We provide two methods to transform an affine LPV system into a parameter-ordered form and reduce the dimension of the parameter space. The first method reduces the complexity by establishing an affine upper bound of the system Gramians which is extended to time-varying parameters. With affine parametric dependence of Gramians a Hankel-norm approximation optimisation is formulated to achieve parameter reduction. The second one is based on considering sensitivity functions of the transfer function and time evolution equations. Both methods are applied to a random system and a thermal model to show the performance of the methods.
For affine linear parameter-varying (LPV) systems, this paper develops two parameter reduction methods for reducing the dimension of the parameter space. The first method achieves the complexity reduction by transforming the affine LPV system into a parameter-ordered form and establishing an affine upper bound of the system Gramians, which is extended to time-varying rate-bounded parameters.  The second method is based on considering the sensitivity function of the transfer function and time evolution equations. Both methods are applied to an academic example and a thermal model. Simulation results together with some analysis are given.
\end{abstract}

\section{Introduction}
    In the past decade, the class of linear parameter-varying (LPV) systems has been developed and established as a reliable and very efficient model class for characterizing nonlinear systems, representing parametric uncertainty and gain scheduling purposes. Many successful applications ranging from very-large-scale integration (VLSI) to aircraft designs\cite{Benner2015,Marcos2004,Dettori2001} have been based on implementations of the LPV framework.
    The inherent complex nature of physical systems often results in high dimensional models with large dimensional state spaces and large dimensional parameter spaces. Typically, the dimension of the parameter space grows as the complexity of the system increases. In practice, it is often necessary to evaluate system performance over substantial ranges of parameter values. We find this theme in problems where design parameters need to be tuned, calibration problems, geometrical optimisation in circuits \cite{Daniel2004} and MEMS devices \cite{Baur2011} and robustness analyses in control systems. To have a reasonable computational complexity in terms of synthesis and simulation, model order reduction for parametrised systems (pMOR) is often required.

    In the aforementioned applications, a high dimensional state space model is often derived from a high resolution spatial discretization of partial differential equations (PDEs). Typically, if high precision is required, this process results in many first order ordinary differential equations (ODEs) approximating the solutions of the PDE. Moreover, the physical constraints and design parameters of such high-fidelity models lead to large dimensional parameter and state spaces. A number of relevant approximation problems can then be phrased as follows:
    \begin{enumerate}[leftmargin=*]
        \item the \emph{state reduction problem} involves the reduction of the dimension of the state space, while preserving accuracy and the physical meaning of the parameters.
        \item the \emph{state and parameter reduction problem}  involves the simultaneous reduction of the dimension of the state space and the dimension of the parameter space.
        \item the \emph{parameter reduction problem} involves the reduction of the dimension of the parameter space only.
    \end{enumerate}

    In most pMOR work, the primary goal has been to solve the \emph{state reduction problem}. A number of such methods have been proposed, mainly focusing on sampling techniques \cite{Panzer2010,Benner2014,Bui-Thanh2008}. On this topic, two influential survey papers \cite{Benner2014,Benner2015} have appeared. Methods based on moment matching of the parametrised transfer function are of particular interest, but tend to become complex with the number of matched moments. For example, in \cite{Benner2014}, the dimension of the reduced model increases exponentially as the number of parameter interpolation points and moments increase. One solution to this problem is to interpolate both the state space and the parameter state within the predefined variation range \cite{lou2018parametric}. Even though it is not always stated, simple sampling schemes imply static dependence on the parameter. Techniques which explicitly deal with time-varying parameters are more involved \cite{Wood1995}. All these works are confined to state reduction without considering the problem to reduce the number of parameters. Approaches which consider parameter reduction either require typical trajectories of the parameter \cite{Rizvi2016}, lack interpretation or are limited in application \cite{Sun2006}. The development of more general parameter reduction techniques can significantly improve the efficiency of simulations, often without loss of generality, as was shown in \cite{Feng2006}. This leads to the second challenging problem: the \emph{state and parameter reduction problem}. In \cite{Feng2006} a two-step approach is introduced. First, parameter reduction is employed to find a low-dimensional parameter space. The second step amounts to constructing a state reduction via moment-matching. However, the reduced rank regression method used in the first step only quantifies the relation between the parameters and the outputs, which is limited by the type of input excitation used. Besides, the system dynamics are not taken into account over ranges of parameters. Given the fact that parameter spaces are usually determined by the physics and the design constraints, it is relevant to explore the correlation between the parameter space and system theoretical properties such as reachability and observability of parametrised systems.

   Our intention, therefore, is to focus on the \emph{parameter reduction problem}. Firstly, we exploit the relationship between the parameter space and system Gramians. A projection-based method is proposed as the means to reduce the dimension of the parameter space using Hankel-norm approximation. Secondly, we give an analysis of the sensitivity of the evolution equations to parameter changes. In doing so, some definitions of system norms for LPV systems are introduced, aimed at characterizing approximation errors in a consistent manner. The methods in this paper are developed for time-invariant parameters with an extension of the Gramian based approach to the time-varying rate-bounded case. % Especially with regard to the generalisation to time-varying parameters, these approaches will be applicable to a wide variety of LPV systems.

    The paper is organised as follows. In Section II we provide a brief introduction to LPV systems and introduce the system norms that will be used in this paper. After a formalisation of the  problem, an argument for affine Gramian reduction
    is presented. Next, the cross-correlation among the parameter space in both frequency domain and time domain
    is developed. Section IV provides a sensitivity analysis of the parameters on the transfer function. In Section V, the results are illustrated in an academic example and in a real application of a thermal model which consists of several interconnected components. Conclusions and future work are presented in Section VI.

    \section{Preliminaries and notation}
    Consider a system $\Sigma(\theta)$ defined by the following LPV state-space representation
    \begin{subequations}    \label{eq:LPVstatespace}
        \begin{align}
            \dot{x}(t) &= A(\theta)x(t) + B(\theta)u(t) \\
            y(t) &= C(\theta)x(t) + D(\theta)u(t)
        \end{align}
    \end{subequations}
where $x\in \mathbb{R}^{n}$ is the state variable, $u\in \mathbb{R} ^{m}$ and $y \in \mathbb{R}^{q}$ denote the input and output, respectively. Furthermore, $\theta$ represents the parameter that is assumed to reside in the parameter space $\Theta\subseteq\mathbb{R}^{\ell}$. A distinction is made between systems where $\theta$ is time-variant and time-invariant. In this work, we consider $\dot{\theta}=0$ unless stated otherwise.
    The state space matrices are assumed to have affine dependence on $\theta$, i.e., the system matrix $A(\theta)$ is given by
    \begin{equation}
    A(\theta) =  A_0+A_1\theta_1 + \cdots + A_{\ell}\theta_{\ell}.
    \end{equation}

    To exploit the parameter dependence of the system, we rewrite the system matrices as linear fractional representations, i.e., $A(\theta)$ is represented as
    \begin{equation}\label{eq:affinenotation}
       A(\theta)
        % = \left(
        %\begin{bmatrix}
        %    1& \theta_1 & \cdots & \theta_{\ell}
        %\end{bmatrix}
        %\otimes I_{n}\right)
        %\begin{bmatrix}
        %    A_0 \\
        %    \vdots \\
        %    A_{\ell}
        %\end{bmatrix}
        %        \\\nonumber
        %        &= A_0 + \mathbf{\theta}_{n_x}A^\theta
        %        \\ \nonumber
        = \bar{\theta}_{n}A^\theta
    \end{equation}
	where
	\begin{equation}
	     \label{tp}
        \bar{\theta}_{n}= [1,\theta_1,  \cdots ,\theta_{\ell}] \otimes I_{n}\in \mathbb{R}^{n \times n(\ell +1)}
     \end{equation}
     and
     \begin{equation}\label{eq:vectorMatrix}
        A^{\theta} = \begin{bmatrix}
            A_0 \\
            \vdots \\
            A_{\ell}
        \end{bmatrix}\in \mathbb{R}^{n(\ell+1) \times n} %, \quad A^{\theta}_i = A_i \text{ for }i = 0,\cdots,\ell
     \end{equation}
       is the matrix of column stacked matrices. All superscripted matrices represent stacked matrices obtained in this manner. Here $\otimes$ denotes tensor multiplication. Similar expressions apply to $B({\theta})$, $C({\theta})$ and $D({\theta})$. Furthermore, the parameter space $\Theta$ is assumed to be the convex hull of $k$ generating vectors $w_j$, $j=1,\ldots, k$, so that $\Theta =\text{Co}\left\{w_1, \dots w_{k}\right\}$ represents the parameter space. Without loss of generality, the system is assumed to be scaled such that $\theta_i \in [0,1], i = 1,\dots, \ell$. It is also assumed that the LPV system is quadratically stable, observable and reachable for all $\theta\in \Theta$.
    %In the presented notation, subscripts $\theta$ represent parameter dependencies while superscripts $\theta$ represent stacked state space matrices. Throughout the paper, the short-hand notation $\mathbf{\bar{\theta}}_{n_\bullet}$ will be used for tensor products of the form \eqref{tp}.
    Using the above notation the continuous time affine LPV system can be rewritten as
    \begin{subequations}\label{eq:affinestatespace}
        \begin{align}
        	    \dot{x}(t) &= \bar{\theta}_{n}A^\theta x(t) + \bar{\theta}_{n}B^\theta u(t) \\
                y(t) &= \bar{\theta}_{n}C^\theta x(t) + \bar{\theta}_{n}D^\theta u(t).
        \end{align}
    \end{subequations}
    The parameter projection methods proposed in this paper are defined as linear transformations $T\in\mathbb{R}^{(\ell+1)\times (\ell+1)}$ of the original parameter space. Specifically, let the transformed parameter $\tilde{\theta}$ be defined as
    \begin{align}        \label{eq:paramtransformation}
        \tilde{\theta} := [1, \theta_1,\cdots,\theta_{\ell}]    \cdot  T
    \end{align}
    where $T$ has full rank and satisfies the orthonormality property %$TT^T=T^TT=I_{(\ell+1)}$.
	 \begin{align}
        TT^T=T^TT=I_{(\ell+1)}.
        \label{eq:orthonormalT}
    \end{align}
    Any such $T$ defines the state space matrices of a transformed system according to
  %   \eqref{eq:affinestatespace}. With $T$, %a system $\tilde{\Sigma}(\theta)$ is defined. the transformed system matrices are transformed according to
    \begin{subequations}        \label{eq:sysmatrixtransform}
        \begin{align}
            {A}(\tilde{\theta}) &=  ( \tilde{\theta}\otimes I_{n})(T^T\otimes I_{n})A^\theta
            \\
            {B}(\tilde{\theta}) &= ( \tilde{\theta}\otimes I_{n})(T^T\otimes I_{n})B^\theta
            \\
            {C}(\tilde{\theta}) &= ( \tilde{\theta}\otimes I_{n})(T^T\otimes I_{n})C^\theta
            \\
            {D}(\tilde{\theta}) &= ( \tilde{\theta}\otimes I_{n})(T^T\otimes I_{n})D^\theta
        \end{align}
    \end{subequations}
    where $(T\otimes I_{n})$ denotes the block-diagonal matrix $\text{diag}(T,\ldots ,T)$. Then it is easily seen that, by %the the orthonormality property,
    (\ref{eq:orthonormalT}),
    the condition %we can obtain the equivalence shown as
\begin{align}\label{eq:equvilenceTrans}
    (\tilde{\theta}\otimes I_n)(T^T\otimes I_n) &= ([1,\theta_1,\cdots,\theta_\ell]\cdot T \otimes I_n)(T^T\otimes I_n)\nonumber \\
    & =([1,\theta_1,\cdots,\theta_\ell]\otimes I_n)(TT^T\otimes I_n)\nonumber \\
    & = [1, \theta_1,\cdots,\theta_\ell] \otimes I_n
    \end{align}
guarantees system equivalence in the sense that
${A}(\tilde{\theta} )= A(\theta$), ${B}(\tilde{\theta}) = B(\theta)$,
${C}(\tilde{\theta} )= C(\theta$), ${D}(\tilde{\theta})= D(\theta)$.

After introducing transformation of the parameter space, a reduction of the parameter space is achieved by choosing
   $T_{r}\in\mathbb{R}^{(\ell+1)\times n_r}$ with rank $n_r < \ell$ and by setting
    \begin{equation}\label{eq:parameterredc}
       \tilde{\theta}_{r} := [1, \theta_1,\cdots,\theta_{\ell}] \cdot T_{r}
     %   \begin{bmatrix}  T_1 & T_2 & \dots & T_{n_r} \end{bmatrix} = \theta
    \end{equation}
    as the lower dimensional parameter vector. In this case, $T_r$ no longer satisfies \eqref{eq:orthonormalT} but defines an orthonormal projection whenever $T_r^TT_r=I_{n_r}$.
    By applying the parameter projection $T_r$ defined above, the transformation \eqref{eq:sysmatrixtransform} results in a system ${\Sigma}(\tilde{\theta}_r)$ with $n_r<\ell$ parameters. The state space matrices of the reduced system ${\Sigma}(\tilde{\theta}_r)$ are
    \begin{subequations}\label{eq:RedSysmatrixtransform}
        \begin{align}
            {A}(\tilde{\theta}_r) &=  ( \tilde{\theta}_r\otimes I_{n})(T_r^T \otimes I_{n})A^\theta
            \\
            {B}(\tilde{\theta}_r) &= ( \tilde{\theta}_r\otimes I_{n})(T_r^T \otimes I_{n})B^\theta
            \\
            {C}(\tilde{\theta}_r) &= ( \tilde{\theta}_r\otimes I_{n})(T_r^T \otimes I_{n})C^\theta
            \\
            {D}(\tilde{\theta}_r) &= ( \tilde{\theta}_r\otimes I_{n})(T_r^T \otimes I_{n})D^\theta.
        \end{align}
    \end{subequations}

\begin{figure}[tbh!]
        \centering
        		\begin{tikzpicture}
                    \draw  (-0.5,0.75) rectangle (0.5,0.25)node[xshift=-0.5cm, yshift=0.25cm]{$1/s$};
                    \draw  (1,1.125) rectangle (-1,2.375)node[xshift=1cm, yshift=-.625cm]{${\Sigma}(\theta )$};
                    %\draw  (-0.5,3.25) rectangle (0.5,2.75)node[xshift=-0.5cm, yshift=0.25cm]{$\tilde{\Delta}_{\theta}$};
                    %\draw  (-1,-1.125) rectangle (1,-2.375)node[xshift=-1cm, yshift=.75cm]{$\tilde{\Sigma}_r$};
                    \draw  (-1,-0.25) rectangle (1,-1.5)node[xshift=-1cm, yshift=.625cm]{${\Sigma}(\tilde{\theta}_r)$};
                    %\draw  (-0.5,-2.75) rectangle (0.5,-3.25)node[xshift=-0.5cm, yshift=0.25cm]{$q$};
                    \draw  (-0.5,-1.875) rectangle (0.5,-2.375)node[xshift=-0.5cm, yshift=0.25cm]{$1/s$};
                    %\draw  (-0.5,-0.25) rectangle (0.5,-0.75)node[xshift=-0.5cm, yshift=0.25cm]{$\tilde{\Delta}_{\theta r}$};

                    \draw[->]  (2.5,0) ellipse (0.25 and 0.25);
                    \node [xshift=2.25cm,yshift=0.5cm]{$+$};
                    \node [xshift=2.25cm,yshift=-0.5cm]{$-$};

                    %\draw[->] (1,-1.5)node[xshift=0.25cm,yshift=0.25cm]{$z_r$} -- (1.5,-1.5) -- (1.5,-0.5) -- (0.5,-0.5);
                    %\draw[->] (-0.5,-0.5) -- (-1.5,-0.5) -- (-1.5,-1.5) -- (-1,-1.5)node[xshift=-0.25cm,yshift=0.25cm]{$w_r$};
                    %\draw[->] (-0.5,-3) -- (-1.5,-3) -- (-1.5,-2) -- (-1,-2)node[xshift=-0.25cm,yshift=-0.25cm]{$x_r$};
                    %\draw[->] (1,-2)node[xshift=0.25cm,yshift=-0.25cm]{$x^+_r$} -- (1.5,-2) -- (1.5,-3) -- (0.5,-3);
                    \draw[->] (-0.5,-2.125) -- (-1.5,-2.125) -- (-1.5,-1.125) -- (-1,-1.125)node[xshift=-0.25cm,yshift=-0.25cm]{$x_r$};
                    \draw[->] (1,-1.125)node[xshift=0.25cm,yshift=-0.25cm]{$\dot{x}_r$} -- (1.5,-1.125) -- (1.5,-2.125) -- (0.5,-2.125);
                    \draw[->] (1,1.5)node[xshift=0.25cm,yshift=-0.25cm]{$\dot{x}$} -- (1.5,1.5) -- (1.5,0.5) -- (0.5,0.5);
                    \draw[->] (-0.5,0.5) -- (-1.5,0.5) -- (-1.5,1.5) -- (-1,1.5)node[xshift=-0.25cm,yshift=-0.25cm]{$x$};
                    %\draw[->] (1,2)node[xshift=0.25cm,yshift=0.25cm]{$z$} -- (1.5,2) -- (1.5,3) -- (0.5,3);
                    %\draw[->] (-0.5,3) -- (-1.5,3) -- (-1.5,2) -- (-1,2)node[xshift=-0.25cm,yshift=0.25cm]{$w$};
                    \draw[->] (-3.5,0) node[xshift=0.25cm,yshift=0.25cm]{$u$} -- (-2.5,0) -- (-2.5,1.75) -- (-1,1.75);
                    %\draw[->]  (-3.5,0) -- (-2.5,0) -- (-2.5,-1.75) -- (-1,-1.75);
                    \draw[->]  (-3.5,0) -- (-2.5,0) -- (-2.5,-0.875) -- (-1,-0.875);
                    %\draw[->] (1,-1.75)node[xshift=0.75cm,yshift=0.15cm]{$y_r$} -- (2.5,-1.75) -- (2.5,-0.25);
                    \draw[->] (1,-0.875)node[xshift=0.75cm,yshift=0.15cm]{$y_r$} -- (2.5,-0.875) -- (2.5,-0.25);
                    \draw[->] (1,1.75)node[xshift=0.75cm,yshift=0.15cm]{$y$} -- (2.5,1.75) -- (2.5,0.25);
                    \draw[->] (2.75,0) -- (3.5,0)node[xshift=-0.25cm,yshift=0.25cm]{$e$} ;

                    \filldraw[dashed, color=blue, fill opacity = 0.1, even odd rule]
                    %(-1.75,1) rectangle (1.75,-1)
                    %(3,2.5) rectangle (-3,-2.5);
                    (-1.75,1) rectangle (1.75,-0.125)
                    (3,2.5) rectangle (-3,-1.625);

                    %\node[xshift=2.5cm,yshift=-2.25cm,color=blue]{\Large$\Sigma_e$};
                    \node[xshift=2.5cm,yshift=-1.25cm,color=blue]{\large$\Sigma_e(\theta)$};
                \end{tikzpicture}
        \caption{Interconnection of the reduced parameter error system, indicated by the blue area.}
        \label{fig:errorLFR}
    \end{figure}
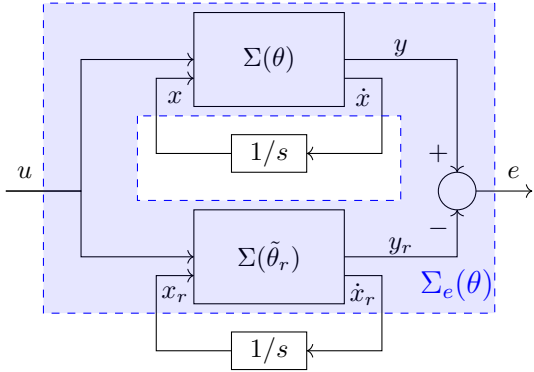

    Noticing that the reduced system matrices have the same rank as the matrices of the original system. To evaluate the performance of the reduced system, an error system is defined as the LPV system with input $u$ and output $y-y_r$ with $y$ and $y_r$ the outputs of (\ref{eq:LPVstatespace}) and \eqref{eq:RedSysmatrixtransform}. See Fig.~\ref{fig:errorLFR}. This system is compactly denoted as $\Sigma_e(\theta)   =  \Sigma(\theta ) - {\Sigma}(\tilde{\theta}_r)$
    %\begin{equation} \label{eq:errorsystem}
    %    \Sigma_e(\theta)   =  \Sigma(\theta ) - {\Sigma}(\tilde{\theta}_r) %\underbrace{\left[\begin{array}{c|c}
        % A(\theta) & B(\theta) \\ \hline
        % C(\theta) & D(\theta)
        % \end{array}\right]}_{{\Sigma}(\theta )}
        % -
        %  \underbrace{\left[\begin{array}{c|c}
        % A(\tilde{\theta}_r) & B(\tilde{\theta}_r) \\ \hline
        % C(\tilde{\theta}_r) & D(\tilde{\theta}_r)
        % \end{array}\right]}_{{\Sigma}(\tilde{\theta}_r)}.
  %  \end{equation}
   where $\tilde{\theta}_r$ is defined in (\ref{eq:parameterredc}). As such, $\Sigma_e(\theta)$ is viewed as an error system in the parameter $\theta$ only. Note that the error system is affinely dependent on the parameters.

     In the LTI setting, the error system can be evaluated by many well established and computable norms. Among these, the $H_\infty$-, $H_2$- and Hankel-norm are commonly used in model reduction. In this work, we introduce a composite error measure on the LPV system that consists of a norm over the parameter space and a system norm
    \begin{equation}\label{eq: pHnorm}
    \left\lVert\Sigma_e(\theta)\right\rVert_{p_{\infty,H}} := \underset{\theta\in\Theta}{\max}\left\lVert\Sigma_e(\theta)\right\rVert_{H},
    \end{equation}
    which evaluates the maximal Hankel-norm of the system $\Sigma(\theta)$ when ranging over the feasible parameter space. Based on what we have discussed so far, we give the problem formulation of parameter reduction for LPV systems.

    \textit{Problem}: Given $\Sigma(\theta)$ such as \eqref{eq:LPVstatespace} and $n_r < \ell$, find $T_r \in \mathbb{R}^{(\ell+1)\times n_r}$ and $\Sigma(\tilde{\theta}_r)$ such that
    \begin{equation}\label{eq:ProblemFormulation}
     \left\lVert\Sigma_e(\theta)\right\rVert_{p_{\infty,H}} =  \underset{\theta\in\Theta}{\max}\left\lVert\Sigma(\theta ) - {\Sigma}(\tilde{\theta}_r)\right\rVert_{H}
    \end{equation} is minimal.
    %In section \ref{sec:affine Gramian}, a detailed discussion of this norm is given. Other norms such the average or square root Hankel-norm are discussed in the Appendix \ref{appendix: LPVnorm}.

    %In the LTI setting, error systems can be evaluated in any of the well established and computable norms. Of these, the $H_\infty$-, $H_2$- and Hankel-norm are commonly used in model reduction.
    %In this paper, we summarise  a notation and definition which may be applied to extend LTI norms to LPV norms

    \section{Hankel-norm reduction}
    It is well known that the Hankel norm of a stable LTI system can be expressed in terms of reachability and obervability Gramians \cite{Antoulas2005}. We establish a similar result for LPV systems first.
    The Hankel operator associated with a stable LTI system $\Sigma$ is defined by
    \begin{gather}
        \mathcal{H}: \mathcal{L}_{2}(\mathbb{R}_-^{m}) \longmapsto \mathcal{L}_{2}(\mathbb{R}_+^{q}),  u_{-} \longmapsto y_{+} \nonumber \\
        \text{where} \quad \mathcal{H}(u_{-}) (t) = \int_{-\infty}^{0} H(t-\tau)u(\tau) d\tau, t \in \mathbb{R}_{+},
        \label{eq: hankeloperator}
    \end{gather}
    it maps the past inputs $u_{-}$ into future outputs $y_{+}$. Here, $H$ is the impulse response of $\Sigma$. The $\mathcal{\ell}_2$-induced norm of $\mathcal{H}$ is defined as
    \begin{gather}
        \left\lVert\mathcal{H}\right\rVert_{\mathcal{L}_2-\text{ind}}  = \sup_{\left\lVert u_-\right\rVert_2} \frac{\left\lVert y_+\right\rVert_2}{\left\lVert u_-\right\rVert_2} = \left\lVert\Sigma\right\rVert_{H}.
    \end{gather}

    The quantity $||\Sigma||_H$ is the Hankel norm of the system $\Sigma$ and
    equals the spectral norm $||\Sigma||_{H}  = \sigma_{\text{max}}(\mathcal{H})$. The Hankel singular values of the system $\Sigma$ are defined as the singular values of the Hankel operator $\mathcal{H}$ associated with $\Sigma$.
    \begin{lemma}
        Given a reachable, observable and stable LTI system $\Sigma$ of dimension $n$, the Hankel singular values are equal to the absolute value of the eigenvalues of the product of $\mathcal{PQ}^T$
        \begin{gather}\label{eq: Hankelsingularvalue}
        \sigma_i(\Sigma) = \sqrt{\lambda_i(\mathcal{PQ}^T)}, i = 1,...,n
        \end{gather}
        where $\mathcal{P}$ and $\mathcal{Q}$ are the controability Gramian and observability Gramian of $\Sigma$.
    \end{lemma}

    For a time-invariant LPV system, the Hankel operator is defined as %for every $\theta\in\Theta$.
    \begin{gather}\label{eq: hankeloperatorparameterised}
        \mathcal{H}_\theta: \mathcal{L}_{2}(\mathbb{R}^{m}_-\times\mathbb{P}) \longmapsto \mathcal{L}_{2}(\mathbb{R}^{q}_+),\  u_{-},\theta \longmapsto y_{+} \nonumber \\
        \text{where}
        \\\nonumber
        \quad \mathcal{H}_\theta(\theta,u_{-})(t) = \int_{-\infty}^{0} H(\theta,t-\tau)u(\tau) d\tau, t \in \mathbb{R}_{+},\theta\in\Theta,
    \end{gather}
    it maps the past inputs and parameters into future outputs. It follows that the $\ell_2$-induced norm of $\mathcal{H}_\theta$ is now parameter dependent. Since $\Sigma(\theta)$ is assumed to be stable $\forall\theta\in\Theta$, the Hankel norm of $\Sigma(\theta)$ can be expressed as
    \begin{gather}\label{eq: hankelnormMaxe}
        \left\lVert
        \Sigma(\theta)
        \right\rVert_{H} = \sigma_{\text{max}}(\mathcal{H}_{\theta}) = \sqrt{\lambda_{\text{max}} \left(\mathcal{P}(\theta)\mathcal{Q}^T(\theta)\right)},
    \end{gather}
    where $\mathcal{P}(\theta),\mathcal{Q}(\theta)$ are the reachability and observability gramians, defined as the unique solutions of the parametrized Lyapunov equations
    \begin{align}
     A(\theta)\mathcal{P}(\theta) + \mathcal{P}(\theta)A^T(\theta) +  B(\theta)B^T(\theta) &= 0\label{eq:lpvlyapunov1} \\
     A(\theta)^T\mathcal{Q}(\theta) + \mathcal{Q}(\theta)A(\theta) +  C(\theta)^TC(\theta) &= 0.\label{eq:lpvlyapunov2}
    \end{align}

     Finding an exact solution to the Lyapunov equation for the whole parameter space is not trivial and often intractable. In the literature, a static Gramian \cite{Scherer2005} is proposed which often suffices, but leads to conservative solutions. For parameter dimension reductions, the parameter dependent Gramians are necessary as they express changes in the system due to parameter variations.  %A static matrix often suffices, as parameter dependence is not a requirement for making statements on stability. Therefore, an affine dependent function is chosen as the candidate.

    The following result shows that a relaxation of (\ref{eq:lpvlyapunov1}) to an inequality naturally leads to an upper bound on $\mathcal{P}(\theta)$ and $\mathcal{Q}(\theta)$ for all $\theta$. Subsequently, an upper bound of the Hankel norm \eqref{eq: hankelnormMaxe} is derived.
    %The resulting Lyapunov inequality is given
%    \vspace{-10pt}
%    \begin{equation}\label{eq:lyapunovequality}
%    \mathbf{\bar{\theta}}_{n_x}\left[\bar{A}^\theta P^T + P\bar{A^\theta}^T + \bar{B^\theta}\bar{B^\theta}^T\right]\mathbf{\bar{\theta}}^T_{n_x} \preceq 0.
%    \end{equation}
    \begin{theorem}\label{thm:upperbound}
        Consider an LPV system (\ref{eq:LPVstatespace}) that is stable, reachable and observable for all $\theta \in \Theta$. There exist unique solutions $\mathcal{P}(\theta)=\mathcal{P}^T(\theta)\succeq0, \mathcal{Q}(\theta)=\mathcal{Q}^T(\theta)\succeq 0$ which satisfies (\ref{eq:lpvlyapunov1}) and (\ref{eq:lpvlyapunov2}).
        Furthermore, there exists $P(\theta)=P^T(\theta)\succeq 0 $ and $Q(\theta)=Q^T(\theta)\succeq 0 $ which satisfy
        \begin{align}
            A(\theta) P(\theta) + P(\theta)A^T(\theta) + B(\theta)B^T(\theta) &\preceq 0. \label{eq:inequLMI1} \\
            A(\theta)^T{Q}(\theta) + {Q}(\theta)A(\theta) +  C(\theta)^TC(\theta)&\preceq  \label{eq:inequLMI2} 0.
        \end{align}
        for all $\theta \in \Theta$. All solutions $P(\theta)$ an $Q(\theta)$ upper bound $\mathcal{P}(\theta)$ and $\mathcal{Q}(\theta)$ in the sense that $P(\theta) \succeq \mathcal{P}(\theta)\succeq 0 $ and $Q(\theta)\succeq \mathcal{Q}(\theta)\succeq 0 $. Moreover,
        the Hankel norm of $\Sigma(\theta)$ is upper bounded by
          \begin{align}\label{eq: Hankelsingularvalueparameterised}
              \lVert\Sigma(\theta) \rVert_H \! = \sigma_{\text{max}}(\mathcal{H}_{\theta}) &=  \!  \sqrt{\!\lambda_{\text{max}} \left(\mathcal{P}(\theta)\mathcal{Q}^T(\theta)\right)}\! \nonumber \\ &\leq \!  \sqrt{\!\lambda_{\text{max}}\left(P(\theta)Q^T(\theta)\right)}.
        \end{align}
        \end{theorem}

        The proof is given in Appendix \ref{app:upperboundproof}. Note that solutions of \eqref{eq:inequLMI1} and \eqref{eq:inequLMI2} are not unique. To measure the importance of the parameter $\theta \in \Theta$ to the system Gramians, a function which is affinely dependent on $\theta$ is of particular interest to find an upper bound of $\mathcal{P}(\theta)$ and $\mathcal{Q}(\theta)$ for all $\theta \in \Theta$.

        %A set $\mathbb{P}$ is defined as all solutions of \eqref{eq:inequLMI1}. Moreover, any element $P(\theta) \subseteq \mathbb{P}$ upper bound $\mathcal{P}(\theta)$ in the sense that $P(\theta)\succeq\mathcal{P}(\theta)\succeq0$. In the same manner, we define a set $\mathbb{Q}$ for all solutions of \eqref{eq:inequLMI2} and $Q(\theta) \subseteq \mathbb{Q}$ upper bound $\mathcal{Q}(\theta)$. As mentioned in the introduction, the error system is chosen affinely dependent on the parameters. Thus, a affinely dependent on $\theta$ function is of particular interest to find an upper bound of $\mathcal{P}(\theta),\theta \in \Theta$. Specifically,
    \begin{theorem}\label{thm:affineFunctionUpperbound}
        Consider an LPV system defined by (\ref{eq:LPVstatespace}) that is stable, reachable and observable for all $\theta \in \Theta$. Suppose that $\Theta = \text{Co}\{w_1,\dots, w_k\}$. Then, for $\mathcal{P}(\theta)$ which satisfies \eqref{eq:lpvlyapunov1} there always exists an affine function $f: \Theta \rightarrow \mathbb{R}^{n \times n}$ such that
        \begin{equation}\label{eq: affineDependentP}
        f(\theta) \succeq \mathcal{P}(\theta) \text{ for all } \theta \in \Theta.
        \end{equation}
        Here $f(\theta)= \mathbf{\bar{\theta}}_{n}P^{\theta}$ and $P^{\theta}$ denote stacked matrices $P(\theta)$ which satisfy \eqref{eq:inequLMI1}.
    \end{theorem}

          \emph{Theorem \ref{thm:affineFunctionUpperbound}} can be proved by making use of the convexity of affine functions $f(\theta)$ with convex hull of $\theta \in \Theta$. The upper bound affine function for observability Gramian is omitted for clarity.
\subsection{Parameter reduction using Hankel-norm approximation}

The main ingredients of parameter reduction for LPV systems are to construct an upper bounded Hankel norm defined in \emph{Theorem \ref{thm:upperbound}} which satisfies the affine form in \emph{Theorem \ref{thm:affineFunctionUpperbound}}, and then to find a reduced model with the least error for the given reduced parameter space. By choosing affine upper bounding Gramians of $\Sigma(\theta)$ and using the transformation introduced previously, the following equivalence can be obtained
    \begin{equation}\label{eq:Equlity1}
      P(\theta)Q^T(\theta) = \bar{\theta}_n  P^{\theta} Q^{\theta^T} \bar{\theta}_n^T.
    \end{equation}

  Following the same arguments \eqref{eq:sysmatrixtransform} as for  $P(\bar{\theta}) = ([1,\theta_1,\cdots,\theta_{\ell}]\cdot T \otimes I_{n})(T^T\otimes I_n)P^{\theta}$, pick up an appropriate $T_{r}\in\mathbb{R}^{(\ell+1)\times n_r}$ with $n_r < \ell$, use the relation in \eqref{eq:equvilenceTrans} and the equivalence is found
    \begin{equation}\label{eq:Equlity2}
     \small{P(\tilde{\theta}_r)Q^T(\tilde{\theta}_r) = \bar{\theta}_n (T_r T_r^T \otimes I_{n})P^{\theta}Q^{\theta^T}(T_rT_r^T\otimes I_{n})\bar{\theta}_{n}^T}.
    \end{equation}

    Once the upper bounded Hankel norm of the original model and the reduced model has been defined and transformed, the problem stated in \emph{Section II} can be reformulated as
    	\begin{equation}\label{eq:OPTIM}
    		\begin{aligned}
    			& \underset{ \tiny{T_{r}}}{\text{min}}   && \underset{\theta \in \Theta}{\text{max}}		\lVert P(\theta)Q^T(\theta)-  P(\tilde{\theta}_r)Q^T(\tilde{\theta}_r) \rVert_H \\
    			& \text{s.t} &&       P(\theta)Q^T(\theta) = \bar{\theta}_n  P^{\theta} Q^{\theta^T} \bar{\theta}_n^T,  \\
			   &      && P^{\theta^T}\mathbf{\bar{\theta}}^T_{n} = \mathbf{\bar{\theta}}_{n}P^{\theta}\succeq 0,\quad Q^{\theta^T} \bar{\theta}_n^T = \bar{\theta}_n Q^{\theta}\succeq 0  \\
			   &&&  T_r \in\mathbb{R}^{(\ell+1)\times n_r}.
    	\end{aligned}
    	\end{equation}

    	Remark that the explicit expression of $P(\tilde{\theta}_r)Q^T(\tilde{\theta}_r)$ is given in  and it is $T_r$ dependent. For solving the above  $\min-\max$ optimisation problem, the first step is to find an upper bound $P(\theta)$ that satisfies \emph{Theorem \ref{thm:upperbound}} and \emph{Theorem \ref{thm:affineFunctionUpperbound}}. This is equivalent to finding a solution of the following inequalities
%The parameter reduction for time-invariant affine LPV systems will be achieved by finding  $f(\theta)$ that satisfies \emph{Theorem \ref{thm:upperbound}} and \emph{Proposition \ref{thm:affineFunctionUpperbound}}. This is equivalent to finding
	\begin{subequations}
        \begin{align}
            \mathbf{\bar{\theta}}_{n}\left[{A}^\theta P^{\theta^T} + P^{\theta} {A^\theta}^T + {B^\theta}{B^\theta}^T\right]\mathbf{\bar{\theta}}^T_{n} \preceq 0&, \\ P^{\theta^T}\mathbf{\bar{\theta}}^T_{n} = \mathbf{\bar{\theta}}_{n}P^{\theta}\succeq 0&.
        \end{align}
        \label{eq:LMIinequality}
    \end{subequations}
    The second step is to construct a $T_r$ which minimises the error between the original system and the reduced one with the maximal $\theta$. This discussion is summarised in \emph{Algorithm \ref{algo:parameterProjection}}. Indeed, the 'optimal' solution of \eqref{eq:OPTIM} is not guaranteed since it is not convex problem.

        \begin{algorithm}[H]
    \caption{: Parameter reduction for affine LPV systems}
    \label{algo:parameterProjection}
    \begin{algorithmic}[1]
    \State{Construct parameter affine dependent upper bounds $\bar{\theta}_nP^{\theta}, \bar{\theta}_nQ^\theta$ that satisfy \emph{Theorem \ref{thm:upperbound}} and \emph{Theorem \ref{thm:affineFunctionUpperbound}}.}
   \State{Obtain $v_{n_r}$ by solving \eqref{eq:OPTIM} for given $n_r$,}
        	\State{Orthonormalise the $v_{n_r}$ such that $v_{n_r}^T v_{n_r} = I_{n_r}$. }
        	\State{Assign $T_{n_r} \leftarrow  v_{n_r}$.}
    \State{\textbf{Return} $ T_{n_r}$}
    \end{algorithmic}
    \end{algorithm}

    The optimum of the quadratic function (\ref{eq:LMIinequality}) of $\theta$ can be evaluated on the vertices of $\Theta$. A formal proof of this observation is given in \cite{Cox2018}. Finding a solution to this Lyapunov inequality is done using linear matrix inequalities (LMIs). Every vertex of the parameter space results in two LMIs \eqref{eq:LMIinequality}. In the case of interval bounding constraints on each entry of the parameter vector, this results in a total of $2^{(\ell+1)}$ LMIs. Like many problems in the LPV setting, this approach becomes intractable with growing dimensions of the parameter space. The difference with other problems is that there is no tuning involved in this process. Therefore there is a one time cost of solving the LMIs.

    %What remains to be proven is the upper bounding property in (\ref{eq:affineGramian}). This can be easily shown since, (\ref{eq:LMI1})$\prec$(\ref{eq:lpvlyapunov1}), which combined with the assumption of quadratic stability leads to the conclusion that $\mathcal{P}(\theta)\preceq P^T\bar{\theta}^T_{n_x}$

%        \begin{subequations}
%            \begin{align}
%            \dot{x} = \mathbf{\bar{p}}_{n_x}\bar{A}^px + \mathbf{\bar{p}}_{n_x}\bar{B}^pu \\
%            y = \mathbf{\bar{p}}_{n_y}\bar{C}^px + \mathbf{\bar{p}}_{n_y}\bar{D}^pu
%            \end{align}
%        \end{subequations}

%    \begin{align}
%        \mathcal{P}(\theta) \preceq P^T\mathbf{\bar{\theta}}^T_{n_x} = \mathbf{\bar{\theta}}_{n_x}P
%        \label{eq:affineGramian}
%    \end{align}

%    The assumption of affine dependency in the Gramian will in specific cases yield an exact solution. One of such cases is where only $A(\theta)$ has affine dependency \todo{proof}.

    The previously presented method assumed static parameter dependence $\dot{\theta}=0$. In the case where the parameter is dynamic but is rate limited $\dot{\underline{\theta}}_i<\dot{\theta}_i<\dot{\bar{\theta}}_i$ for every $\theta_i \subseteq {\theta} $, the Lyapunov inequality may be adjusted according to (\ref{eq:timevariantlyapunov}). The resulting LMIs are again quadratic in $\theta$, now with a constant offset. Therefore it can still be evaluated on the vertices of the parameter space. The formal proof is again found in \cite{Cox2018}.
    \begin{subequations}
        \begin{gather}
            \dot{\mathcal{P}}(\theta) + {A}(\theta)\mathcal{P}(\theta) + \mathcal{P}(\theta)A(\theta)^T + {B}(\theta){B}(\theta)^T \preceq 0
            \\
            \textcolor{white}{\small Affine Gramian}\Downarrow \text{\small Affine Gramian}\nonumber
            \\
            P^{\theta}\dot{\theta} + \mathbf{\bar{\theta}}_{n}\left[{A}^\theta P^{\theta^T} + P^{\theta} {A^\theta}^T + {B^\theta}{B^\theta}^T\right]\mathbf{\bar{\theta}}^T_{n} \preceq 0
            \\
            \textcolor{white}{\small Bounded parameter velocity}\Downarrow \text{\small Bounded parameter velocity}\nonumber
            \\
            \resizebox{.87\hsize}{!}{$L(\theta,\dot{\underline{\theta}}) = P^{\theta}\mathbf{\dot{\underline{\theta}}}_{n} + \mathbf{\bar{\theta}}_{n}\left[{A}^\theta P^{\theta^T} + P^{\theta} {A^\theta}^T + {B^\theta} {B^\theta}^T\right]\mathbf{\bar{\theta}}^T_{n} \preceq0$} \label{eq:lyapunovvelocitylower}
            \\
            \resizebox{.87\hsize}{!}{$L(\theta,\dot{\overline{\theta}}) = P^{\theta}\mathbf{\dot{\bar{\theta}}}_{n} + \mathbf{\bar{\theta}}_{n}\left[{A}^\theta P^{\theta^T} + P^{\theta} {A^\theta}^T + {B^\theta}{B^\theta}^T\right]\mathbf{\bar{\theta}}^T_{n} \preceq0$} \label{eq:lyapunovvelocityupper}
        \end{gather}
     %   \label{eq:timevariantlyapunov}
    \end{subequations}

    Extending the approach to parameters which are rate bounded  increases the number of LMIs to $3^{\ell+1}$. By imposing an additional constraint $P_i^{\theta}\succ 0 $ (We abuse the notation $ P_i^{\theta}$ for $i${th} stacked matrix of $P^{\theta}$ which is referred to the notation in \eqref{eq:vectorMatrix}.) results in $L(\theta,\dot{\underline{\theta}})\prec L(\theta,\dot{\overline{\theta}}) $ and thus (\ref{eq:lyapunovvelocitylower}) can be dropped. This reduces the number of LMIs to $2^{\ell+1}+ \ell$. The previously presented methods to obtain an affine reachability Gramian can be applied for finding an affine observability Gramian $Q(\theta)$ as well.

    In LTI approximation the loss function is related to the energy of the error system since the Hankel singular values are a measure of energy in each state respectively. For the proposed method, the interpretation is more nuanced as the affine Gramians constitute an upper bound on the actual Gramians. If this upper bound is tight, the loss function is a good indication of the error. The tightness of this upper bound is heavily dependent on the system, and determining the tightness is not trivial. Inspired by Adamjan-Arov-Krein (AAK) theorem, we provide a relative $p_{\infty,\infty}$-error which asses the approximation error
    %Instead of evaluating the Hankel-norm over the system, checking the  %The insight into the $p_{\infty,\mathcal{H}}$-norm error determined from the affine Gramian approximation is therefore questionable.
    \begin{align}
        \frac
        {\lVert {\Sigma}(\theta )-\tilde{\Sigma}_r(\theta_r)\rVert_{p_{\infty,\infty}}}
        {\lVert {\Sigma}(\theta )\rVert_{p_{\infty,\infty}}}.
    \label{eq:relativepinferror}
    \end{align}

	Here $||\cdot||p_{\infty,\infty}$ is defined as
	\begin{equation}
		\| \Sigma_{e}(\theta)\|_{p_{\infty, \infty}} := \underset{\theta \in \Theta}{\max}\|\Sigma_{e}(\theta)\|_{H_{\infty}}
	\end{equation}
	which evaluates the $H_\infty$-norm of the system $\Sigma(\theta)$ when varying over the feasible parameter space.

    \section{Sensitivity analysis}
    Another method for parameter reduction is sensitivity analysis, which is similar to principal component analysis \cite{Kwiatkowski2005,Kwiatkowski2005a}. This method evaluates how sensitive the outputs are to small changes in the parameter values. It can be viewed as principal component analysis without the requirement of needing a typical parameter trajectory. Evaluation can be done in either the time or the frequency domain.
    \subsection{Frequency domain}
    Since it is assumed that $\dot{\theta}=0$ the notion of transfer functions is still applicable. The transfer function of an affine LPV system is given below
        \begin{align}\label{eq:transferfunciton}
        H(\theta,q) = D_\theta + C_\theta q(I-A_\theta q)^{-1}B_\theta.
    \end{align}
    The \emph{transfer function sensitivity} is defined as the Jacobian of the transfer function with respect to $\theta$, as shown in the following
    \begin{align}
        {_\theta}H(\theta,q) &= \frac{dD_\theta}{d\theta} +\colorbox{blue!10}{\large$\frac{dC_\theta}{d\theta}q(I-A_\theta q)^{-1}B_\theta$} \label{eq:transfersensitivity}\\
        & + \colorbox{green!10}{\large$C_\theta q\frac{d(I-A_\theta q)^{-1}}{d\theta}B_\theta$}  + \colorbox{red!10}{\large$C_\theta q(I-A_\theta q)^{-1}\frac{dB_\theta}{d\theta}$} \nonumber
     \end{align}

    The $i^{th}$ element of the Jacobian can be represented as the system given in (\ref{eq:jacobianelement}), where the colours correspond to separate elements of the Jacobian as in (\ref{eq:transfersensitivity}). Note  that the resulting system again has affine dependency on the parameters.
    \begin{align}
        \resizebox{.89\hsize}{!}{$
            {^i_\theta}H(q, \theta) =
            \left[\begin{array}{cccc|ccc}
            \bc A_\theta & \bc 0   		& \bc 0   	   & \bc 0   	  & \bc B_\theta \\
            \bc 0   	 & \gc A_\theta & \gc A_i 	   & \gc 0   	  & \gc 0\\
            \bc 0   	 & \gc 0   		& \gc A_\theta & \gc 0   	  & \gc B_\theta  \\
            \bc 0   	 & \gc 0   		& \gc 0   	   & \rc A_\theta & \rc B_i\\ \hline
            \bc C_i 	 & \gc C_\theta & \gc 0   	   & \rc C_\theta & D_i
            \end{array}\right]
            =
            \left[\begin{array}{c|c}
            \mathcal{A}_{\theta,i} & \mathcal{B}_{\theta,i} \\ \hline
            \mathcal{C}_{\theta,i} & D_i
            \end{array}\right]$}
        \label{eq:jacobianelement}
    \end{align}

    From the Jacobian an ordering of the parameters can be determined from the $p_{\infty,\infty}$-norm of each respective element. However, the transfer function sensitivity  does not take into account the cross-correlations between parameters. Multiplying the transfer function sensitivity by its complex conjugate will result in a transfer function sensitivity covariance matrix (TSCM). The elements of this matrix are defined as
    \begin{align}
        \Pi_{ij} := \left\lVert{{^i_\theta}H(\theta,q)}^{*}{{_\theta^j}H(\theta,q)} \right\rVert_{p_{\infty,\infty}}
        \label{eq:TSCM}.
    \end{align}

    This product of systems gives element $i,j$ of the TSCM, in the notation of (\ref{eq:jacobianelement}), and is equal to
    \begin{align}
        \Pi_{i,j} =
        \left\lVert\left[\begin{array}{cc|c}
        \mathcal{A}^*_{\theta,i} & \mathcal{C}^*_{\theta,i}\mathcal{C}_{\theta,j} & \mathcal{C}^*_{\theta,i}D_j \\
        0 & \mathcal{A}_{\theta,j} & \mathcal{B}_{\theta,j} \\\hline
        \mathcal{B}^*_{\theta,i} & D^*_i\mathcal{C}_{\theta,j} & D_i^*D_j
        \end{array}\right]\right\rVert_{p_{\infty,\infty}}
    \end{align}

    All elements of this system have affine dependency on $\theta$, except for $\mathcal{C}^*_{\theta,i}\mathcal{C}_{\theta,j}$ which in general does not possess such a property. Still the $p_{\infty,\infty}$-norm can be evaluated over a finite set of points. This can be shown by extending the parameter space with all quadratic elements of $\theta$. The system will then have affine dependence on the extended parameter space and thus can be evaluated over an extended convex hull. If the extended convex hull is not properly chosen it may lead to conservatism. %As an example take $\theta_1 \in [0\ 1], \theta_2\in [0\ 1]$. The quadratic elements of this parameter space are $\theta_1^2, \theta_1\theta_2, \theta_2^2$. The range of these elements are $\theta_1^2\in[0\ 1]$, $\theta_1\theta_2\in [0\ 1]$ and $\theta^2\in[0\ 1]$. If the convex hull is taken directly from these ranges, the vertex $[0\ 1\ 1]$ would be valid. However the actual system cannot reach such a point since $\theta_1=0$ therefore $\theta_1\theta_2\neq1$.

    The resulting TSCM is a symmetric $n_\theta\times n_\theta$ matrix. By taking the singular value decomposition (SVD) of this matrix, $\Pi=TST^*$, the parameter transformation matrix is found. This transformation orders the parameter directions in terms of transfer sensitivity covariance and is orthonormal, constituting a valid transformation as defined in (\ref{eq:paramtransformation}).

    \subsection{Time domain}
    Unlike principal component analysis, the proposed time-domain approach does not require a typical trajectory of the parameters. Instead the solution is written as time dependent equation. In discrete time the output evolution equation is given as
    \begin{align}
        \resizebox{.89\hsize}{!}{$
            y(k) = C_\theta A_\theta^kx_0 + \left[\sum_{i=1}^{k}C_\theta A_\theta^{i-1}B_\theta u(k-i)\right] + D_\theta u(k)
            $}
    \end{align}

    The sensitivity function at time $k$ is simply calculated as the Jacobian of this equation towards $\theta$. For affine parameter dependency, (\ref{eq:timesensitivity}) can be calculated in terms of the stacked matrices $\bullet^\theta$ and state space matrices $\bullet_\theta$.
    \begin{align}
        \label{eq:timesensitivity}
        {_\theta}y(k) =& \frac{dC_\theta A_\theta}{d\theta}x_0 + \frac{dD_\theta}{d\theta}u(k) + \\ \nonumber
        & \left[\sum_{i=1}^{k}\frac{dC_\theta A_\theta^{i-1}B_\theta}{d\theta}u(k-i)\right]
%                     &=& \frac{dC_p}{dp}A_px_0 + C_p\frac{dA_p}{dp}x_0 + \frac{dD_p}{dp}u(k) + \\ \nonumber
%                      && \left[\sum_{i=1}^{k} \left(\frac{dC_p}{dp}A_p^{i-1}B_p+C_p\frac{A_p^{i-1}}{dp}B_p + C_pA_p^{i-1}\frac{B_p}{dp}\right)u(k-i)\right]
    \end{align}

    In discrete time the sensitivity can be written as a row vector multiplied by a column vector stacking all inputs. Stacking all outputs, ${_\theta}y_0,\ {_\theta}y_1\ \dots\ {_\theta}y_{k_{\max}}$, into a column vector gives a matrix multiplied by the input vector ${^i_\theta}Y={^i}M(\theta)U$. In the particular case where $A_\theta=A_0$ and either $C_\theta=C_0$ or $B_\theta=B_0$, the matrix ${^i}M(\theta)$ is parameter independent. To take cross correlation into account ${^i_\theta}Y$ is pre multiplied by ${^j_\theta}Y^*$, being equal to ${^j_\theta}Y^*{^i_\theta}Y=U^*{^j}M^*{^i}MU$. The largest singular values of ${^j}M^*{^i}M$ result in the peak gain between the sensitivity functions of the $i^{th}$ and $j^{th}$ parameter. The collection matrix of all the maximum singular values $\bar{\sigma}_{ij}$ gives an $n_p\times n_p$ matrix, the sensitivity covariance matrix (SCM).
    \begin{align}
        \bar{S} =
        \begin{bmatrix}
        \bar{\sigma}_{11} && \bar{\sigma}_{12} && \dots && \bar{\sigma}_{1n_\theta}  \\
        \bar{\sigma}_{21} && \bar{\sigma}_{22} && \dots && \bar{\sigma}_{2n_\theta}  \\
        \vdots && \vdots && \ddots & \vdots \\
        \bar{\sigma}_{n_\theta 1} && \bar{\sigma}_{n_\theta 2} && \dots && \bar{\sigma}_{n_\theta n_\theta}
        \end{bmatrix}
    \end{align}

    From the SCM a similar approach is taken to derive the transformation matrix as with the TSCM. One note on this approach is the choice of $k_{\max}$ which should be chosen at least as large as the largest time constant of the system. Choosing $k_{\max}$ small will exclude dynamics and may lead to bad approximations.

    \section{Results}

    To illustrate these methods of parameter reduction through sensitivity analysis and Hankel-norm approximation, two examples systems are used. The first system is an illustrated LPV system and the second a real thermal system.

    \subsection{Illustrative example}
\begin{figure}[tbh!]
        \centering
        \includegraphics[width=\linewidth]{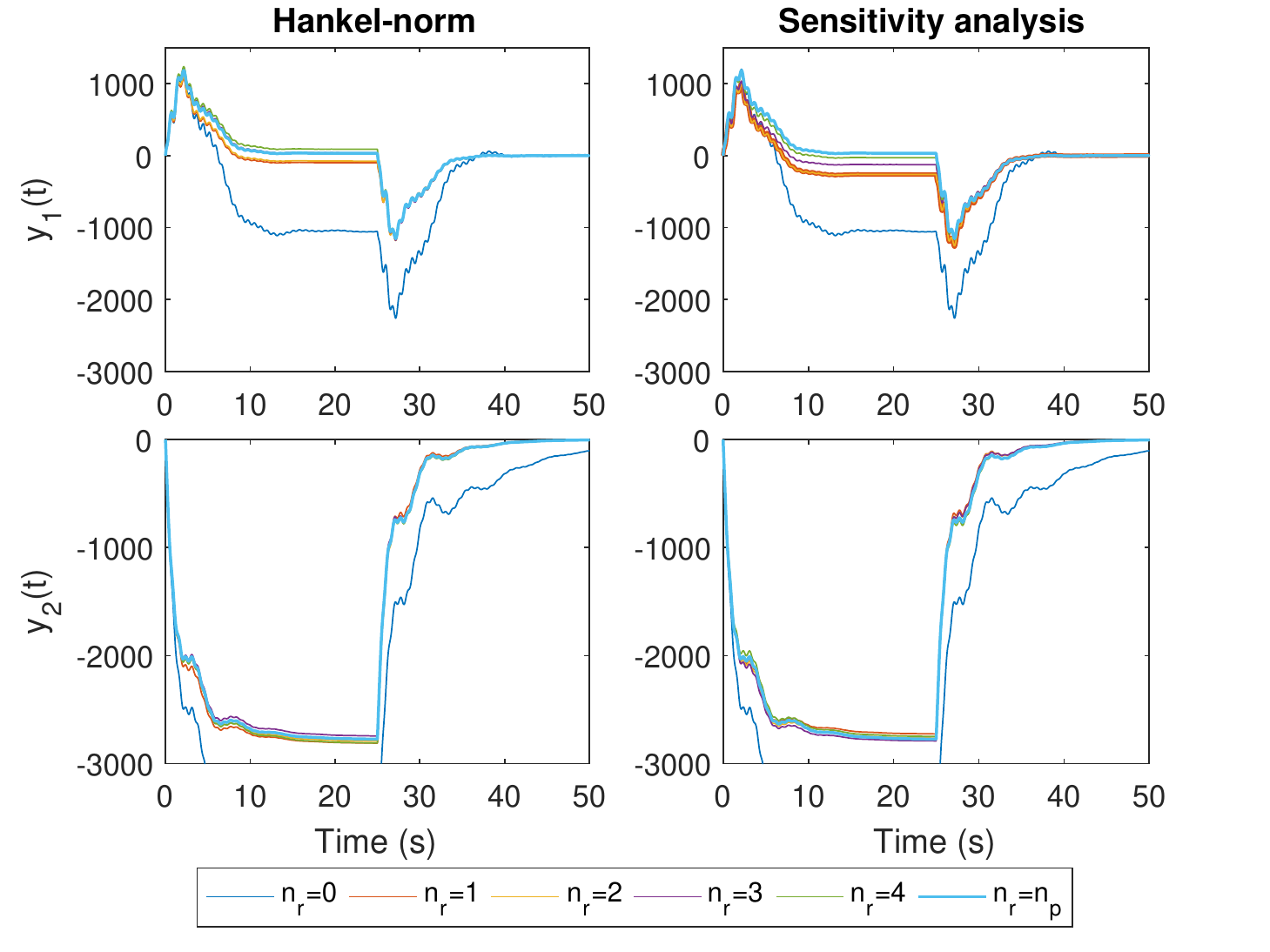}
        \caption{System evolution of the randomised LPV system, at a random parameter realisation, for different reduction orders and methods.}
        \label{fig:random-system-simulation}
        \vspace{-10pt}
    \end{figure}

\begin{figure}[tbh!]
        \centering
        \includegraphics[width=\linewidth]{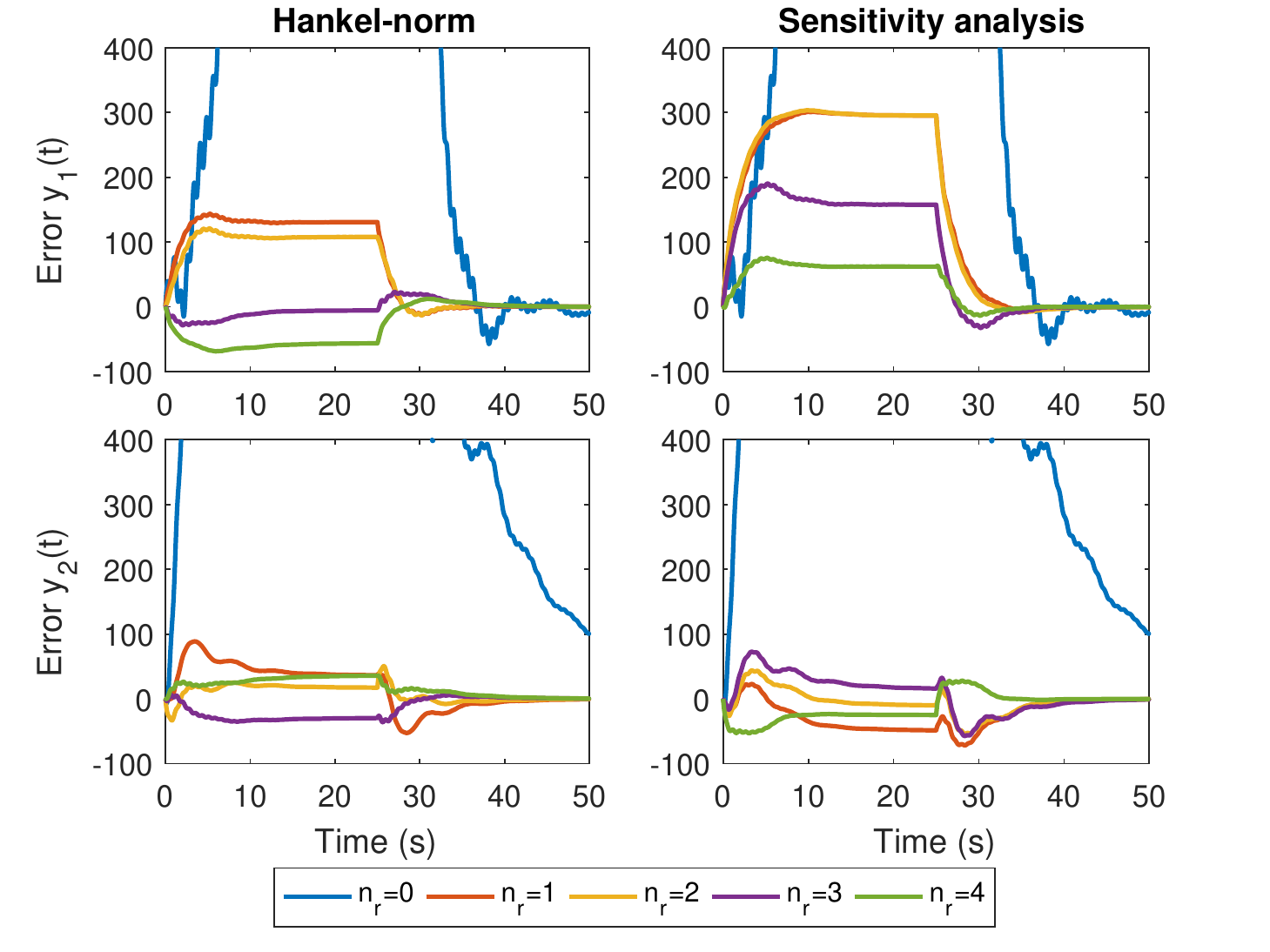}
        \caption{Output error of the randomised LPV system, at a random parameter realisation, for different reduction orders and methods.}
        \label{fig:random-system-simerror}
        \vspace{-10pt}
    \end{figure}

\begin{figure}[tbh!]
        \centering
        \includegraphics[width=\linewidth]{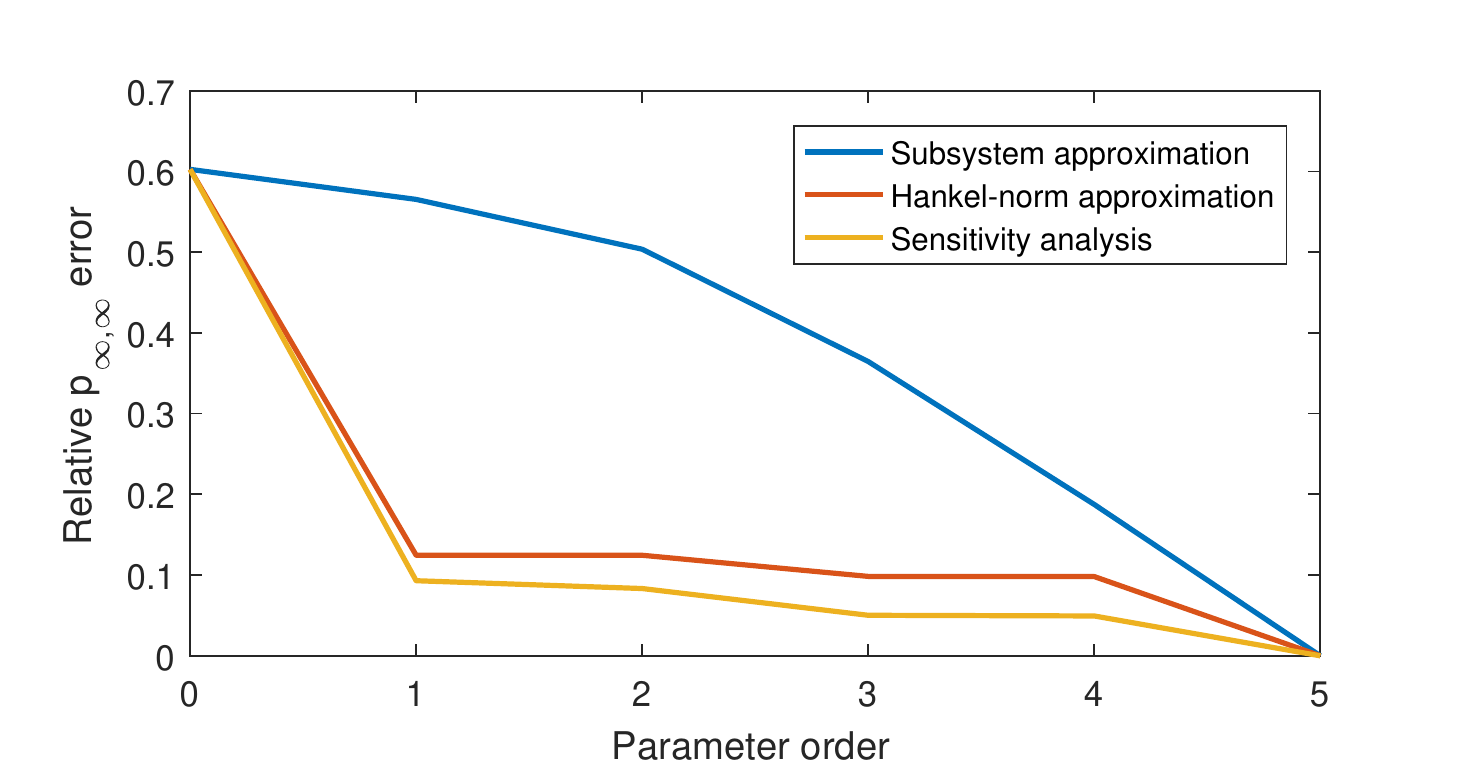}
        \caption{Relative $p_{\infty,\infty}$-error of the randomised LPV system for different parameter orders and approximation methods.}
        \label{fig:random-system-error}
        \vspace{-10pt}
    \end{figure}

    Consider an affine LPV system
    \begin{subequations}
        \begin{align}
            \dot{x} &=  A_0x+B_0u + \sum_{i=1}^5\left(A_i\theta_ix + B_i\theta_iu\right) \\
            y &= C_0x+D_0u + \sum_{i=1}^5\left(C_i\theta_ix + D_i\theta_iu\right),
        \end{align}
        \label{eq: Example1}
    \end{subequations}
    where $\theta_i\in[0,\ 1], u\in\mathbb{R}^2$, $y\in\mathbb{R}^2$ and $x\in\mathbb{R}^{45}$. To ensure stability all $A_i\prec0$. For this system Hankel-norm and TSCM approximations are determined. As a comparison subsystem Hankel-norm approximation is used where reduction is based singular values of actual Hankel norm associated with the remaining parameter space. The $n_p-n_r$ parameters having the smallest 'subsystem' Hankel-norm are removed. In Fig. \ref{fig:random-system-simulation} the output evolution which consists of the transient response and the steady-state response is shown for an input $u(t)=$ constant $\neq 0$ for $t = [0,25]$ and $u(t)= 0$ otherwise. To illustrate the accuracy of the reduced models, the errors are shown in Fig. \ref{fig:random-system-simerror}. For $t = [0,25]$, both figures show a tendency that
    more parameters preserved, the less error is between the reduced and the original model. For $t > 25[s]$, all reduced models with $n_r = 1,...,4$ convergence to the same point. It will change for a different parameter realisation and therefore Fig. \ref{fig:random-system-error} shows the relative $p_{\infty}, \infty$ error of the reductions to be non-increasing with growing parameter order for both methods. It also shows that for this example, the sensitivity analysis outperforms the Gramian based method.


%    \newpage
    \subsection{Thermal simulation}
\begin{figure}[tbh!]
        \centering
        \includegraphics[width=\linewidth]{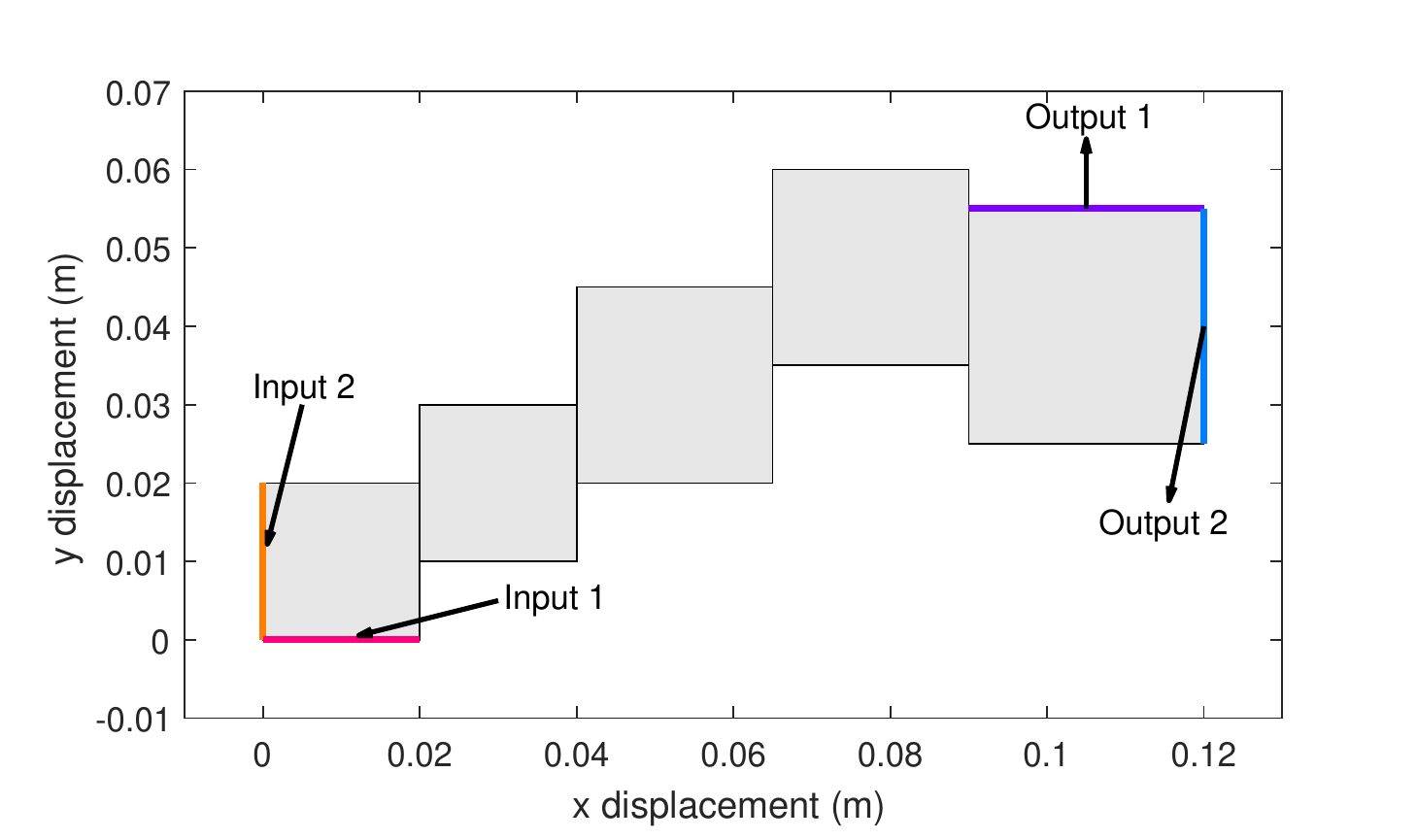}
        \caption{Thermal model with five different material blocks. Orange and pink are heat inputs. Purple and blue are outputs.}
        \label{fig:thermal-system}
        \vspace{-10pt}
    \end{figure}

    The second system is a thermal simulation consisting of five coupled metal blocks, all having parameter dependent heat capacity. Using COMSOL the system is generated with the following structure.
    \begin{subequations}
        \begin{align}
            \dot{x} &= A_0x + B_0u + \sum_{i=1}^{5}\left(A_i\theta_ix +B_i\theta_iu\right) \\
            y &= C_0x
        \end{align}
        \label{eq:thermalstatespace}
    \end{subequations}
    Where $u\in\mathbb{R}^2$, $y\in\mathbb{R}^2$ and $x\in\mathbb{R}^{45}$. The inputs and outputs represent heat power in $[W]$ and temperature $[K]$ respectively. In Fig. \ref{fig:thermal-system} an illustration of the system is shown.

\begin{figure}[tbh]
        \centering
        \includegraphics[width=\linewidth]{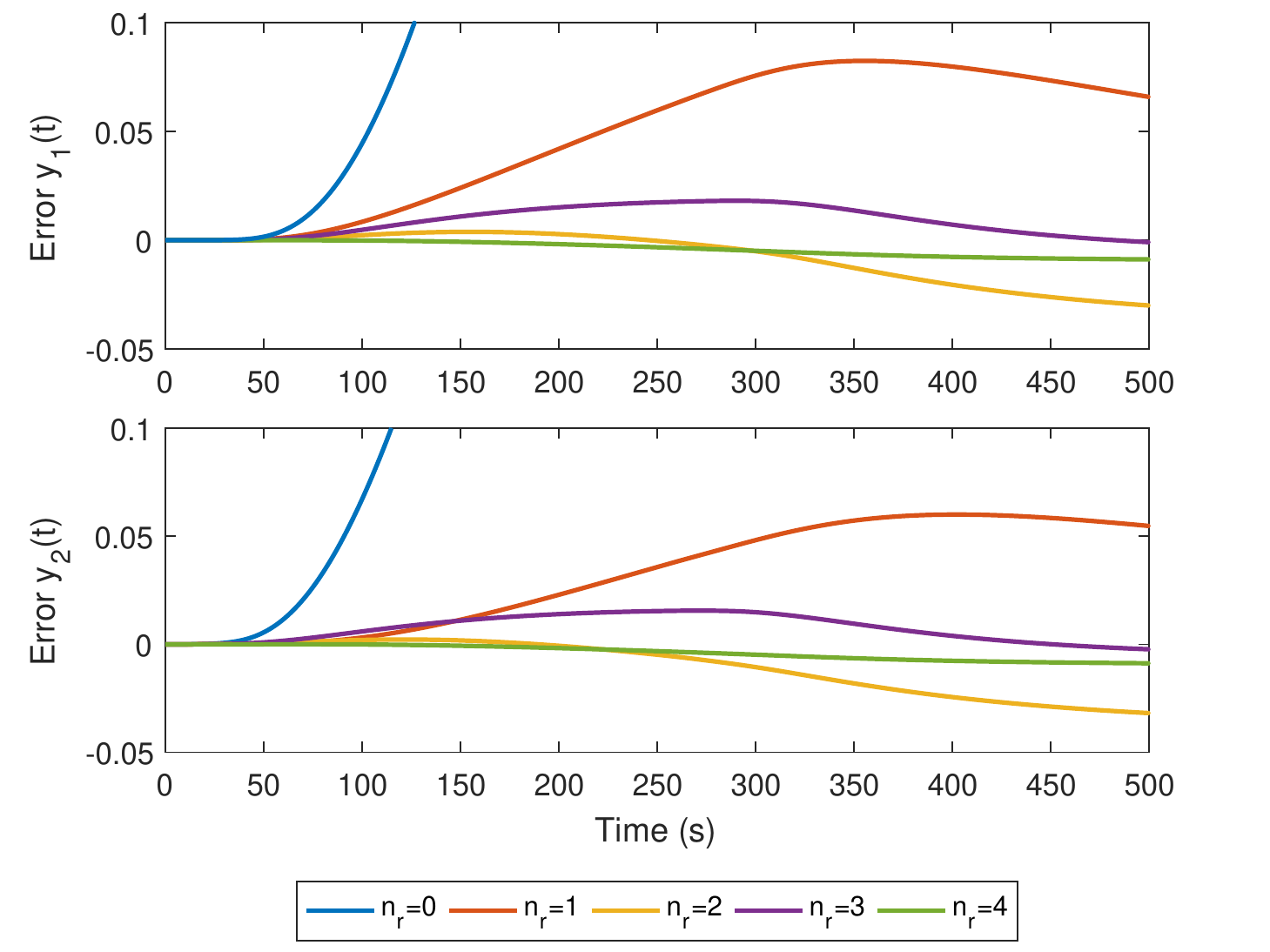}
        \caption{Simulation error of the thermal system for different parameter orders at a randomly selected parameter in the parameter space.}
        \label{fig:reduction-simulation}
        \vspace{-10pt}
    \end{figure}

\begin{figure}[tbh]
        \centering
        \includegraphics[width=\linewidth]{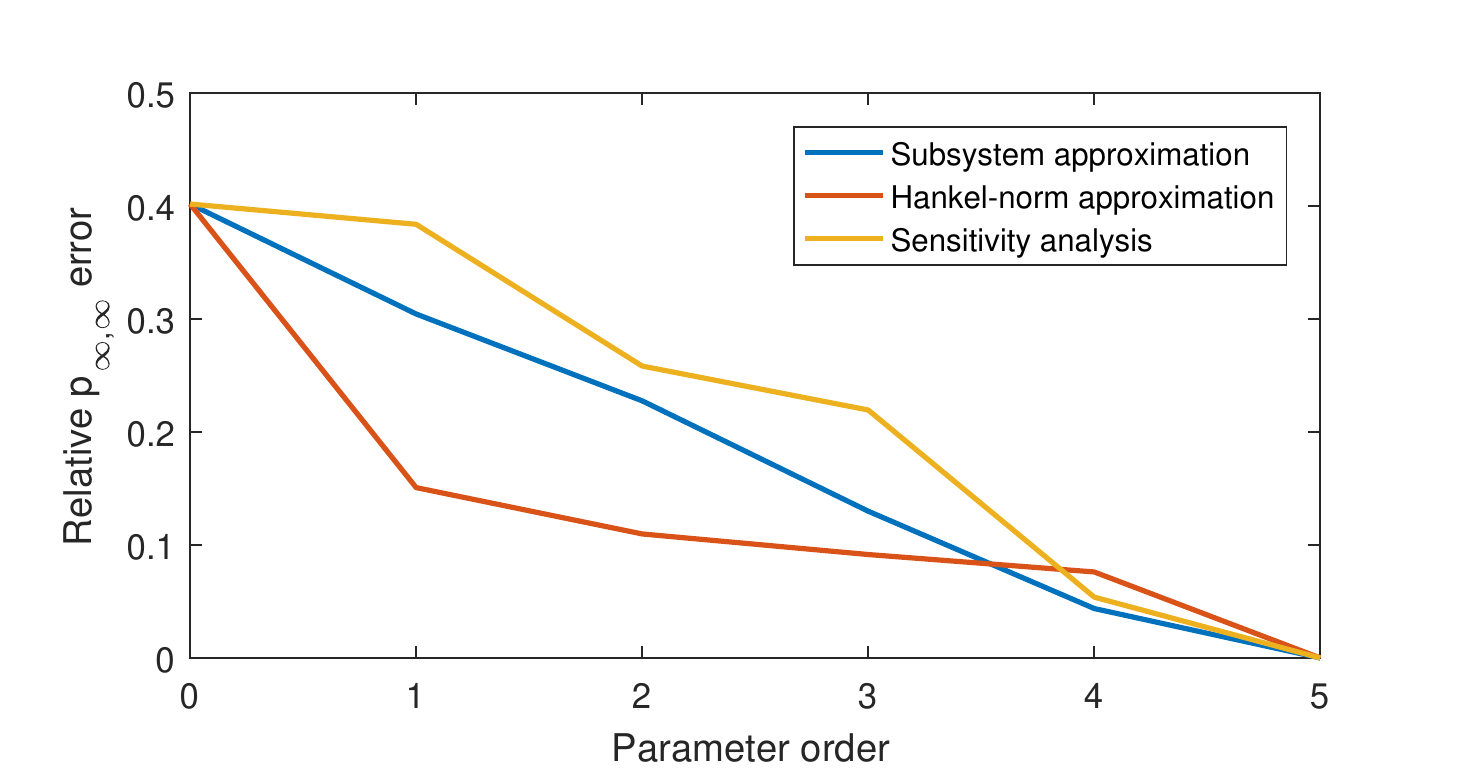}
        \caption{Relative $p_{\infty},\infty$ error of the thermal system for different parameter orders and different reduction methods.}
        \label{fig:reduction-error}
        \vspace{-10pt}
    \end{figure}

    To evaluate the error of the reduced model, a simulation is performed with constant power from $u(t) = [50;45] [W]$ for $ t\in[0\ 250]$ and $u(t) = [0;0]$ afterwards. For different realisations of the parameter values, the error between the reduced model and full order model is shown in Fig. \ref{fig:reduction-simulation}. Clearly the non-parametric model $n_r=0$ performs the worst showing that the nominal model is not a good approximation of the system. The best performing model is the $n_r=4$ model as the error is almost zero. This result shows the performance with respect to a specific parameter space. Therefore, a global error bound is used to infer conclusions on performance.
    In Fig. \ref{fig:reduction-error} the relative $p_{\infty,\infty}$ error is plotted for different reduction orders. This figure illustrates that reducing the system using sensitivity analysis is comparable to reduction in subsystem using Hankel-norm. It is also clear that approximation in the $p_{\infty, H}$ norm using transformation optimisation yields improved results for $n_r<4$. For approximation of $n_r=4$ the Hankel-norm optimisation method performs worse in $p_{\infty},\infty $ error. This is due to a combination of issues, the non-convex optimisation \eqref{eq:OPTIM} and the error introduced in finding the upper bound of the affine Gramian \eqref{eq: affineDependentP}.
%    \todo{maybe remove monte carlo part}

    For both systems computation of the Gramians takes in the order of approximately $1000$ seconds(with a dual core PC and using YALMIP \cite{Lofberg2004}).
    %As a practical note, it should be taken care of to have $A_0,\ B_0$ and $C_0$ of similar magnitude to avoid numerical trouble.
    Because the optimisation in (\ref{eq:OPTIM}) has not shown to be convex, it is ran using different initial condition to find a close approximation of the Hankel singular values. Due to this non-convexity, the resulting transformation matrix does not guarantee the global optimal approximation.

    \section{Conclusions and future work}
     In this paper, two methods for parameter reduction have been given for time-invariant LPV systems. The first method approximates the system using the Hankel-norm over the parameter space. The second method uses principal component analysis on the covariance matrix. A system norm analysis of the performance of these methods has been presented in the paper together with simulation results. It can be concluded that both methods, though substantially different, are computationally feasible and provide good approximations over the parameter space.
	
	Two reduction methods have been presented which focus on reducing the parameter space. The proposed methods are not limited to be only used for simplifying the complexity of parameters, but also can be integrated with state reduction problem as state and parameter reduction problem.

    \bibliographystyle{ieeetr}
    \bibliography{Parameter-reduction-paper}

    \appendices
    \section{Proof of \emph{Theorem 1}}\label{app:upperboundproof}
    %\begin{proof}
    %To show Lyapunov the inequality results in an upper bound of the solution to the equality we start with LTI systems.
    This proof consists of two parts. In the first part, we prove the Lyapunov inequality as an upper bound to the Lyapunov equality. The second part we show that upper bound on the Hankel-Singular value from the upper bounded Gramian which is derived from the first part.

    \subsection{Proof of the Lyapunov inequality as an upper bound to the Lyapunov equality}
    Consider a stable LTI system with stable matrix $A$. Lyapunov equation theorem shows that there is a unique solution $P\succ0$ to
    \begin{gather}
        A\mathcal{P}+\mathcal{P}A^T + Q = 0
    \end{gather}

    where $Q=BB^T\succ 0$. The same system admits a solution space $P\in\tilde{\mathcal{P}}$, being all solutions that satisfy
    \begin{gather}
        AP + PA^T + Q \preceq 0
    \end{gather}

    The equality can be rewritten and substituted into the inequality to give
    \begin{gather}
        A(P-\mathcal{P}) + (P-\mathcal{P})A^T  \preceq 0
    \end{gather}

    From Lyapunov equation theorem and stable $A$, we have $P-\mathcal{P}\succeq 0$ as the solution of the above inequality. Extending to stable, observable and reachable time-invariant LPV systems, it suffices to show that for every $\theta\in\Theta$ a solution can be found to the Lyapunov (in)equality concluding the proof.
    \qedsymbol

        \subsection{Proof of the upper bound on the Hankel-singular values from the upper bounding Gramians}
  %\begin{proof}
        From \cite{bhatia2013matrix} we take two properties of eigenvalue arithmetic of symmetric matrices $A, B \in \mathbb{R}^{n_x \times n_x}$,
        \begin{equation}
                    \begin{aligned} \label{eq:eigenproperties2}
%\label{eq:eigenproperties}
                \lambda_i(A+B)\geq0 & \text{ if } \lambda_i(A)\geq0 \text{ and } \lambda_i(B)\geq0,\ %i=1,\dots,n
                \\
                \lambda_i(A B)\geq0 & \text{ if } \lambda_i(A)\geq0 \text{ and } \lambda_i(B)\geq0,%\ i=1,\dots,n
            \end{aligned}
        \end{equation}
        for $i = 1,...,n$.
       For clarity we drop $\theta$, and define a function $F$ below:
        \begin{align}
                F & :=  PQ - \mathcal{PQ},\notag\\
                &=(P-\mathcal{P})(Q-\mathcal{Q})
                    + \mathcal{P}(Q-\mathcal{Q})
                    + (P-\mathcal{P})\mathcal{Q}. \label{eq:goalequation}
        \end{align}
        Given  $P(\theta)\succeq\mathcal{P}(\theta)\succ 0$ and  $Q(\theta)\succeq\mathcal{Q}(\theta)\succ 0$,
        applying properties (\ref{eq:eigenproperties2}) and \emph{Theorem 1} to (\ref{eq:goalequation}) yields  $\lambda_i(F)\geq 0$, $i=1,\dots,n$.
%        \begin{equation}\label{eq:positiveEigenvalue}
%        \lambda_i(F)\geq 0 \Rightarrow \lambda_i(PQ-\mathcal{PQ})\geq 0 ,\quad i=1,\dots,n.
%        \end{equation}
        With the properties of Hankel matrix $PQ = Q^{\frac{1}{2}}PQ^{\frac{1}{2}}$, for the symmetric $F = (Q^{1/2}PQ^{1/2} - \mathcal{Q}^{1/2}P\mathcal{Q}^{1/2}) = F^T$ it is proven. Therefore, $ PQ\succeq\mathcal{PQ} $ is concluded.
        Next consider the eigenvector $x_{1}$ associated to the largest eigenvalue of $\mathcal{PQ}$. Then the following holds
        \begin{equation}\label{eq:proofEigenvlueHankel}
        x^T_{1}\mathcal{PQ}x_{1} \leq x^T_{1}(PQ)x_{1}  \Rightarrow \lambda_{1}(\mathcal{PQ}) \leq \frac{ x^T_{1}(PQ)x_{1} }{||x_{1}||_{2}^{2}}.
        \end{equation}

		By the definition of eigenvalue decomposition, we have
		\begin{equation}
			\frac{ x^T_{1}(PQ)x_{1} }{||x_{1}||_{2}^{2}} \leq \sup_{y_1} \frac{ y^T_{1}(PQ)y_{1} }{||y_{1}||_{2}^{2}} = \lambda_{\text{max}}(PQ),
		\end{equation}
		here $y_1$ is the eigenvector associated to the largest eigenvalue of $PQ$. Thus, the $\lambda_{\text{max}}(\mathcal{PQ})\leq \lambda_{\text{max}}(PQ)$ is proved. \qedsymbol
\end{document}